\documentclass{aa501}
\usepackage{graphicx}
\begin{document}
   \title{Search for nearby stars among proper motion stars 
          selected by optical-to-infrared photometry}

   \subtitle{I. Discovery of LHS~2090 at spectroscopic distance of
             $d \sim$ 6~pc}

   \author{R.-D. Scholz
          \inst{1} 
          \and
          H. Meusinger
          \inst{2}\fnmsep\thanks{Visiting astronomer, German-Spanish 
Astronomical Centre, Calar Alto, operated by the Max-Planck-Institute for 
Astronomy, Heidelberg, jointly with the Spanish National Comission for 
Astronomy}
          \and
          H. Jahrei{\ss}
          \inst{3}
          }

   \offprints{R.-D. Scholz}

   \institute{Astrophysikalisches Institut Potsdam, An der Sternwarte 16,
              D--14482 Potsdam, Germany\\
              \email{rdscholz@aip.de}
         \and
             Th\"uringer Landessternwarte Tautenburg, 
             D--07778 Tautenburg, Germany\\
             \email{meus@obelix.tls-tautenburg.de}
         \and
             Astronomisches Rechen-Institut, M\"onchhofstra{\ss}e 12-14, 
             D--69120 Heidelberg, Germany\\
             \email{hartmut@ari.uni-heidelberg.de}
             }

   \date{Received ; accepted }

   \abstract{
We present the discovery of a previously unknown very nearby star
- LHS~2090 at a distance of only $d=6$~pc. In order to find nearby 
(i.e. $d < 25$~pc) red dwarfs, we re-identified high proper motion
stars ($\mu >$~0.18~arcsec/yr) from the NLTT catalogue 
(Luyten \cite{luyten7980}) in 
optical Digitized Sky Survey data for two different epochs and in the 
2MASS data base. Only proper motion stars with large $R-K_s$ colour index 
and with relatively bright infrared magnitudes ($K_s<10$) were selected
for follow-up spectroscopy. The low-resolution spectrum of LHS~2090
and its large proper motion (0.79~arcsec/yr) classify this star as 
an M6.5 dwarf. The resulting spectroscopic 
distance estimate from comparing the infrared $JHK_s$
magnitudes of LHS~2090 with absolute magnitudes of M6.5 dwarfs is 
$6.0\pm1.1$~pc assuming an uncertainty in 
absolute magnitude of $\pm$0.4 mag.
   }

   \titlerunning{Search for nearby stars among proper motion stars. 
                 I. Discovery of LHS~2090 at $d \sim$6~pc}
   \maketitle
%
%________________________________________________________________

\section{Introduction}

Our knowledge on the stellar content of the solar neighbourhood is still very 
incomplete. From the statistics of the Catalogue of Nearby Stars one can infer
that at 10 pc about 30\% of all stars are so far undetected (Henry et al.
\cite{henry97}). But the detailed observation of the very nearby stars
is one of the main
starting points for investigations of the stellar luminosity function,
the initial mass function as well as for the search for planetary systems.
Future missions for the detection of extrasolar planets (SIM, TPF, DARWIN)
will concentrate on very nearby stars ($d < 10$~pc) in order
to be able to reveal not only Jupiter-class but also Earth-like planets.

High proper motion catalogues, e.g. the Luyten Half Second (LHS) catalogue
(Luyten \cite{luyten79}) and the New Luyten Two Tenths (NLTT) catalogue
(Luyten \cite{luyten7980}) contain most of the known nearby stars
($d < 25$~pc). All 58 stars in the Catalogue of Nearby Stars (CNS3)
of Gliese \& Jahrei{\ss} (\cite{gliese91}) with distances less than 5~pc
have proper motions larger than 0.5~arcsec/yr. Among the 280 stars within
10~pc contained in the CNS3, there is only one M dwarf with a proper motion
below the NLTT limit of 0.18~arcsec/yr. In the current CNS4 (not yet published)
all stars within 10~pc do have proper motions above that limit.

The majority of the $\sim$60000 NLTT stars have not yet been investigated 
further in order to determine their distances. Among the faint and red NLTT
stars one can expect to find about 40\% to lie within 25~pc (Jahrei{\ss} et al.
\cite{jahreiss01}). The $\sim$3500 LHS stars received 
much more attention in spectroscopic and photometric follow-up observations.  
Nevertheless, the discovery of nearby stars among the LHS stars is still going
on (Henry et al \cite{henry97}; Gizis \& Reid \cite{gizis97}; 
Jahrei{\ss} et al. \cite{jahreiss01}). 

One reason for the incompleteness of the catalogue of nearby stars is the
lower limiting magnitude of the proper motion catalogues in the southern
sky. The northern sky was covered by the Palomar Observatory Sky Survey (POSS)  
observations starting in the 50s, and these Schmidt plates constitute the
first epoch of 
Luyten's high proper motion star surveys (e.g. Luyten \cite{luyten79}).
Nevertheless, few faint high proper motion stars can be found from second
epoch POSS observations (Monet et al. \cite{monet00}). South of $\delta = 
-30^{\circ}$, which is the POSS survey limit in the southern sky,
only recently deep high proper motion surveys have been started using
Schmidt plates (Scholz et al. \cite{scholz00}; Ruiz et al. \cite{ruiz01}).
New high proper motion stars as bright as the active M5 star APMPM~J0237-5928
($R=13.4$, spectroscopic distance: 12~pc) discovered by Scholz et al. 
(\cite{scholz99}) can be found in that region.

There are several attempts to detect nearby stars neglecting the
proper motion information during the first steps of the search. 
One possibility is to search for extremely red faint objects, i.e.
very late-type M dwarfs and the new class of L dwarfs obtained in the 
Two Micron All Sky Survey (2MASS) 
(Reid et al. \cite{reid00}; Gizis et al. \cite{gizis00}) and in the
DEep Near-Infrared Survey (DENIS)  (Delfosse et al. \cite{delfosse01}). 
Another way to detect missing nearby M dwarfs consists in the subsequent 
observation of X-ray sources in order to
identify young M dwarfs with small space motions not present in proper motion 
catalogues (Fleming \cite{fleming98}). But all new nearby stars found in these
surveys turned out to be high proper motion stars, at least in those cases when
distances of less than about 15~pc were measured. 

As a logical consequence, we have started a search for the missing stars in
the solar neighbourhood by combining the proper motion catalogues with 
near-infrared and optical sky surveys. The 2MASS data base (2nd incremental
release public data base) and the A2.0 catalogue (Monet et al. \cite{monet98}) 
can well be used for that purpose. The proper motion stars, however, have to be 
re-identified in digitized sky survey (DSS) images at two different epochs.

%__________________________________________________________________

%
%________________________________________________________________

\section{Combining proper motion and photometry}

Samples of very red (optical-to-infrared colour) point sources in
the 2MASS or DENIS survey may contain compact extragalactic sources,
distant red giants, nearby red dwarfs and very nearby asteroids. If the
source is relatively faint in the optical ($R>10$) and shows a proper 
motion of the order of 0.1 to 10~arcsec/yr (with Barnards's star still being 
the record holder in stellar proper motions), it must be a dwarf. For a
red giant with that faint apparent magnitude, the proper motion would transform
to a very large space velocity ($>1000$~km/s), which is much larger than the 
Galactic escape speed (cf. Leonard \& Tremaine \cite{leonard90}; Meillon et
al.~\cite{meillon97}).

In a previous paper (Jahrei{\ss} et al. \cite{jahreiss01}) we selected red
NLTT stars (Luyten's spectral classes ``m'' and ``m$+$'') as candidates of 
nearby stars and obtained more accurate positions and photographic photometry
from the APM sky catalogues (Irwin et al. \cite{irwin94}). However, the
optical colours of these late-type star candidates show only a weak correlation 
with spectral subclass (Scholz et al. \cite{scholz00}). As already mentioned
in Jahrei{\ss} et al. (\cite{jahreiss01}), optical-to-infrared colours are
a much better choice. The 2MASS data base 
allows an all-sky search with the input of target lists or by pre-defined 
search parameters.

We have cross-identified NLTT stars in a region covering about 50\% of 
the northern sky ($07^{\mathrm{h}} < \alpha < 19^{\mathrm{h}}, 
\delta > 0^{\circ}$)
with the 2MASS data base. Only bright 2MASS sources ($K_s<10$)
identified with faint (Luyten's $R>14.5$) NLTT stars within a search radius of
60~arcsec were further investigated. All identifications were checked with two
epochs of DSS images. In addition to the optical magnitude
estimates given in the proper motion catalogue, we also extracted the
$R$ magnitudes from the A2.0 catalogue (Monet et al. \cite{monet98}).
The DSS1 images usually had the same epoch as the A2.0 catalogue, whereas
the DSS2 images had an epoch close to that of the 2MASS data.

LHS~2090 turned out as a candidate nearby star in our first sample of about
50 stars selected for follow-up classification spectroscopy.
With $K_s=8.4$ and $R-K_s=6.4$ (USNO-2MASS), LHS~2090 was one of
the most promising objects in the list of candidates. The whole sample of
about 35 objects which was successfully observed spectroscopically, will
be subject of a forthcoming paper.

%__________________________________________________________________

%
%________________________________________________________________

\section{Former information on LHS~2090}

LHS\,2090 was detected by Willem Luyten within the scope of his proper motion
survey with the 48-inch Schmidt telescope and got the detection designation
LP\,368-128. Its refined (LHS) proper motion is $0.785''$ in $216.3^{\circ}$.
Willem Luyten's estimates of its red and photographic magnitudes are $R = 15.5$ 
and $m_{\mathrm{pg}} = 17.4$, respectively. 
Luyten hoped that the mean error of the
magnitudes are not larger than $\pm$0.6 mag {\it though errors of even 1.5
do occur} (Luyten \cite{luyten64}). Having this in mind Luyten's 
magnitudes are in good agreement with the USNO-A2.0 values $R = 14.8$ and 
$B = 17.7$, which are also measurements of the POSS-I O and E plates. 
A finding chart of LHS\,2090 was published in the LHS Atlas (Luyten \& Albers
\cite{luyten79b}). Luyten attributed a spectral class ``m'' to this object. 
Since Luyten's publication, no additional information on LHS\,2090 became 
available unless the most recent astrometric and photometric measurements in 
USNO-A2.0 and 2MASS ($09^{\mathrm{h}} 00^{\mathrm{m}} 23\fs59, 
+21\degr 50\arcmin 05\farcs5$ (J2000 at 2MASS epoch 1998.44)). 
It was too faint to enter the photometric surveys of LHS 
stars carried out by Eggen or Weis.

%__________________________________________________________________

%
%________________________________________________________________

\section{Spectroscopic distance estimate of LHS~2090}

Spectroscopic follow-up observations were carried out with CAFOS, the
focal reducer and faint objects spectrograph at the 2.2m Calar Alto telescope.
The grism B-400 was used yielding 9.6\AA\, per pixel on the SITe1d CCD and
a wavelength coverage from 4000\AA\, to 8000\AA. A slit width of 1~arcsec
was used, corresponding to a spectral resolution of 18\AA. The exposure time 
was 60 seconds for LHS~2090 and 120 seconds for the bright comparison star
GJ~1111 (observed during twilight).

The spectra were calibrated by standard procedures within the MIDAS package 
for bias-subtraction, flat-fielding and wavelength calibration using Ne-Ar lamp
spectra. For relative flux calibration we have used spectra of secondary 
spectrophotometric standard stars from Oke \& Gunn (\cite{oke83}).

%
%                                                One column figure
%----------------------------------------------------------- 6pcstar
   \begin{figure}
   \centering
  \includegraphics[angle=-90,width=9cm]{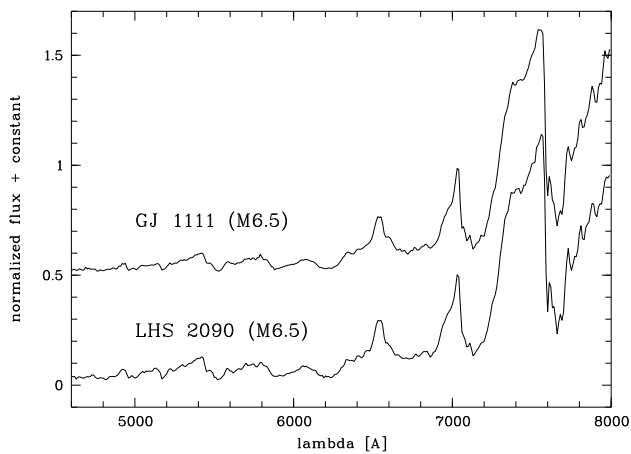}
  \caption{Spectrum of LHS~2090 in comparison to that of the well known
           M6.5 star GJ~1111. Both
           spectra were taken with the 2.2m telescope at Calar Alto.
           The spectra are nearly identical, and we adopted the spectral
           type of M6.5 for LHS~2090.
      }
     \label{FigSpec}
   \end{figure}
%
%______________________________________________________________

The low-resolution spectrum
of LHS~2090 is nearly identical to that of the known M6.5 dwarf GJ~1111 
(see Figure~\ref{FigSpec}). With
that spectral type and taking the mean absolute magnitudes of M6.5
dwarfs $M_J=10.51, M_H=9.94, M_{K_s}=9.60$ given in Kirkpatrick
\& McCarthy (\cite{kirkpatrick94}), we obtained spectroscopic
distance estimates $d_J=6.06$~pc, $d_H=6.07$~pc, $d_{K_s}=5.83$~pc,
respectively. A conservative assumtion of $\pm$0.4 mag accuracy in 
absolute magnitude yields $6.0\pm1.1$~pc.

%__________________________________________________________________

%
%______________________________________________________________

\section{Conclusions}

   \begin{enumerate}
      \item After the recent discovery of DENIS-P~J104817.7-395606.1 (Delfosse
et al. \cite{delfosse01}), an M9 dwarf at only 4~pc spectroscopic distance 
from the Sun, we have found another very nearby star, LHS~2090, at 6~pc.
      \item Despite its large proper motion ($\mu=0.785$~arcsec/yr), LHS~2090
had not been further investigated up to now.
      \item The low-dispersion spectrum of LHS~2090 closely resembles that 
of the M6.5 star GJ~1111. We conclude therefore that LHS~2090 
is of spectral 
type M6.5 as well. Hence the distance estimate based on the 2MASS $JHK_s$ 
magnitudes yields $d=6$~pc. Trigonometric parallax determination and 
investigation of a possible binarity of LHS~2090 is needed for confirmation.
      \item The combination of high proper motion star data with infrared
sky surveys such as the 2MASS survey is a very effective tool for finding 
previously unknown nearby red dwarfs.
   \end{enumerate}

\begin{acknowledgements}
The spectroscopic confirmation of nearby star candidates is based on 
observations made with the 
2.2\,m telescope of the German-Spanish Astronomical Centre, Calar Alto, Spain,
This research has made use of the SIMBAD database and the
VizieR Catalogue Service, Strasbourg, of the STScI Digitized Sky Survey
and of data products from the Two Micron All Sky Survey, which
is a joint project of the University of Massachusetts and the Infrared
Processing and Analysis Center, funded by the National Aeronautics and
Space Administration and the National Science Foundation.

We would like to thank Darja Golikowa for her help with the re-identification
of high proper motion stars in DSS images and their cross-correlation with 
USNO A2.0 and 2MASS data.

RDS gratefully acknowledges financial support from the Deutsches Zentrum
f\"ur Luft- und Raumfahrt (DLR) (F\"or\-der\-kenn\-zeichen 50~OI~0001).

We thank the referee, John Gizis, for his prompt report.
\end{acknowledgements}


\begin{thebibliography}{}

   \bibitem[2001]{delfosse01} Delfosse X., Forveille T., 
      Martin E.L. et al. 2001, 
      A\&A, 366, L13

   \bibitem[1998]{fleming98} Fleming T.A. 1998, 
      ApJ, 504, 461

   \bibitem[1997]{gizis97} Gizis J.E., Reid I.N. 1997,
      PASP, 109, 849

   \bibitem[2000]{gizis00} Gizis J.E., Monet D.G., Reid I.N., Kirkpatrick J.D.,
      Liebert J., Williams R.J. 2000,
      AJ, 120, 1085

   \bibitem[1991]{gliese91} Gliese W., Jahrei{\ss} H. 1991, 
      Preliminary Version of the Third Catalogue of Nearby Stars, 
      computer-readable version on ADC Selected Astronomical Catalogs Vol.1 - 
      CD-ROM

   \bibitem[1997]{henry97} Henry T.J., Ianna P.A., Kirkpatrick J.D., 
      Jahrei{\ss} H. 1997,
      AJ, 114, 388

   \bibitem[1994]{irwin94} Irwin M.J., Maddox S.J., McMahon R.G. 1994, 
      Spectrum N$^o$. 2, 14-16

   \bibitem[2001]{jahreiss01} Jahrei{\ss} H., Scholz R.-D., Meusinger H., 
      Lehmann I. 2001,
      A\&A, 370, 967

   \bibitem[1994]{kirkpatrick94} Kirkpatrick J.D., McCarthy D.W. 1994,
      AJ, 107, 333

   \bibitem[1990]{leonard90} Leonard P.J.T., Tremaine S. 1990,
      ApJ, 353, 486

   \bibitem[1964]{luyten64} Luyten W.J. 1964,
      Proper motion survey with the 48 inch Schmidt telescope.
      Vol II The North Pole.
      University of Minnesota, Minneapolis

   \bibitem[1979]{luyten79} Luyten W.J. 1979,
      LHS Catalogue. Second Edition. University of Minnesota, Minneapolis

   \bibitem[1979-1980]{luyten7980} Luyten W.J. 1979-1980,
      New Luyten Catalogue of Stars with Proper Motions Larger than Two Tenths
      of an Arcsecond, University of Minnesota, Minneapolis, computer-readable
      version on ADC Selected Astronomical Catalogs Vol.1 - CD-ROM

   \bibitem[1979]{luyten79b}  Luyten W.J., Albers H. 1979,
      LHS atlas. An atlas of identification charts for LHS stars.
      University of Minnesota, Minneapolis

   \bibitem[1997]{meillon97} Meillon L., Crifo F., Gomez A., Udry S., 
     Mayor M. 1997,
     in Battrick B. (ed.), Hipparcos Venice'97, ESA-SP~402, p.~591

   \bibitem[1998]{monet98} Monet D., Bird A., Canzian B. et al.  1998,
     USNO-A V2.0, A Catalog of Astrometric Standards,
     U.S. Naval Observatory Flagstaff Station (USNOFS) and
     Universities Space Research Association (USRA) stationed at USNOFS

   \bibitem[2000]{monet00} Monet D.G., Fisher M.D., Liebert J., Canzian B., 
     Harris H.C., Reid I.N. 2000,
     AJ, 120, 1541

   \bibitem[1983]{oke83} Oke J.B., Gunn J.E. 1983,
     ApJ, 266, 713

   \bibitem[2000]{reid00} Reid I.N., Kirkpatrick J.D., Gizis J.E., 
     Dahn C.C., Monet D.G., Williams R.J., Liebert J., Burgasser, A.J. 2000,
     AJ, 119, 369

   \bibitem[2001]{ruiz01} Ruiz M.T., Wischnjewsky M., Rojo P.M., 
     Gonzalez L.E. 2001,
     ApJSS, 133, 119

   \bibitem[1999]{scholz99} Scholz R.-D., Irwin M., Schweitzer A., 
     Ibata R. 1999,
     A\&A, 345, L55

   \bibitem[2000]{scholz00} Scholz R.-D., Irwin M., Ibata R., Jahrei{\ss} H., 
     Malkov O.Yu. 2000,
     A\&A, 353, 958

\end{thebibliography}
\end{document}